\documentclass[a4paper]{jpconf}
\usepackage{iopams}
\usepackage{graphicx,
longtable}

\begin{document}
%

\title{\centerline{Beta Decay and the Cosmic Neutrino Background} }
%
\author{Amand~Faessler}
\address{Institute f\"{u}r Theoretische Physik der Universit\"{a}t
T\"{u}bingen,\\
                D-72076 T\"{u}bingen, Germany; e-mail: faessler@uni-tuebingen.de;}
\author{Rastislav~Hodak}
\address{Institute of Experimental and Applied Physics, Czech Tecnical University, 
Horsk\'a 3a/22, 12800 Prague, Czech Republic.} 

\author{Sergey Kovalenko}

\address{Universidad T\'ecnica Federico Santa Mar\'\i a, \\
Centro-Cient\'\i fico-Tecnol\'{o}gico de Valpara\'\i so, \\
Casilla 110-V, Valpara\'\i so,  Chile}

\author{Fedor ~\v{S}imkovic} 
\address{JINR, 141980 Dubna, Moscow Region,  Russia  \\ and Comenius University, Physics Dept., SK-842 15 Bratislava, Slovakia.}  

\begin{abstract}

In 1964 Penzias and Wilson detected the Cosmic Microwave Background (CMB). 
Its spectrum follows Planck's black body radiation formula and shows a remarkable constant temperature 
of $ T_{0\gamma}   \approx  2.7 $ K independent of the direction. 
The present photon density is about $ 370 $ photons per $ cm^3 $. 
The size of the hot spots, which deviates only in the fifth decimal of the temperature from the average value, tells us, that the universe is flat. 
About 380 000 years after the Big Bang at a temperature of $ T_{0\gamma} =  3000$ K already in the matter dominated era the electrons 
combine with the protons and the $^4He $ and the photons move freely in the neutral universe. So the temperature and distribution of 
the photons give us information of the universe 380 000 years after the Big Bang. 
Information about earlier times 
can, in  principle, 
be derived from the Cosmic Neutrino Background $ (C\nu B) $. The neutrinos decouple already 
1 second after the Big Bang at a temperature of 
about  $10^{10}$ K.
Today their temperature 
is $ \sim1.95$ K 
and the average density is 56 electron-neutrinos per $ cm^3 $.   Registration of these neutrinos is an extremely  
challenging experimental problem which can hardly be solved with the present technologies.  On the other hand it 
represents a tempting opportunity to check one of the key element of the Big Bang cosmology and to probe the early 
stages of the  universe evolution.
The search for the $ C \nu B$ with the induced beta decay 
$ \nu_e + ^3H  \rightarrow  ^3He + e^-$ is the topic of this contribution.
The  signal would show up by a peak in the electron spectrum with an energy of  the neutrino mass above the Q value. 
We discuss the prospects of this approach and argue that it is able to set limits on the $ C \nu B$ density in our vicinity.
We also discuss critically ways to increase with 
modifications of the present KATRIN spectrometer the source intensity by a factor $100$, which would yield about $ 170 $ counts of 
relic neutrino captures per year. This would make the detection of the Cosmic Neutrino Background possible. Presently such 
an increase seems not to be possible.  But one should be able to find an
upper limit for the local density of the relic neutrinos (Cosmic Neutrino
Background) in our galaxy.

\end{abstract}
\newpage


\section{Introduction}

\vspace{0.5cm}

The Cosmic Microwave Background (CMB) was detected in 1964 by Penzias and Wilson $ ~\cite{Penzias} $  (Nobel Prize 1978) 
as byproduct of their seach for possible perturbations of the communication with satellites. It supported the Big Bang (BB) model for the early universe. 
The temperature followed exactly Planck's distribution and was surprisingly identical to four digits independent of the direction ($ T_{0\gamma} = 2.7255(6) $ Kelvin). 
This led to the `inflationary' model. The satellite observations COBE (Cosmic Background Explorer) $ ~\cite{COBE} $ in 1990, 
WMAP (Wilkinson Microwave Anisotropy Probe) $~\cite{WMAP}$ 2001 and Planck $ ~\cite{Planck} $  in 2013 did show deviations of the constant temperature 
of the background photons in the fifth digit. The size of these `hot spots'  
indicated a flat universe. These measurements show a picture of our universe at the decoupling of the photons from matter and give us information 
of the universe about 380 000 years after the big bang (BB) at a temperature of about 3000 Kelvin and the scale of the expanding universe roughly $ 1/1000 $ of today. 
The radius of this sphere of the last interaction is from each observer in the universe about 13 to 14 billion light years (ly) away, reflecting the age of the universe. 
\vspace{0.5cm}

Therefore with photons we are unable to probe the universe  closer  than 300 000 years  to the BB.
The relative abundance of 
light nuclear isotopes allows indirectly to look to events minutes after the BB. 
A similar probe like the photons are the neutrinos. Due to their weak interaction they decouple earlier about 
1 second after the BB from matter at a temperature of $ T_{0\nu} \approx 1 MeV \approx
10^{10} \:  Kelvin$. The Cosmic Neutrino Background $ (C \nu B) $ of today $ T_{0\nu} \approx 1.95$ K contains 
therefore information of the universe already $\sim$1 second after the BB. 
The detection of this 
$ C \nu B $ seems to be hardly possible due to the weak interaction of neutrinos with matter and due to their low energy 
$ ( T_\nu = 1.95$ K  $\approx 2\cdot 10^{-4} [eV]) $. Nevertheless several methods have been 
discussed in the literature to search for these relic neutrinos ~\cite{Faessler4,Lazauskas, Ringwald, Wigmans, Ringwald2, 
Weiler, Vega, Munyaneza, Paes1}. 
Let us mention among them the following proposals:

\begin{itemize}

\item Annihilation of an ultrahigh energy cosmic neutrino with a relic antineutrino into the Z-resonance $ ~\cite{Paes1} $. Energy and momentum conservation require, 
that the Z-boson has an energy of  $ E_Z = 4 \cdot 10^{22} \: [eV] $. If the decay Z-burst is directed to the earth one should observe several gamma rays, 
nucleons or electrons above the Greisen-Zatsekin-Kuzmin cutoff of about $ 10^{21} [eV] $, if the event is within our GZK-radius off about $ 50 Mpc $. 

\item Like in the case of dark matter the solar system with the earth is probably moving with some speed through the relic neutrino cloud $ ~\cite{Hagmann} $. 
One fixes a cylindrical object free in space frictionless with magnetic forces. Divided by  a plane going through the cylindrical symmetry axis half the material 
can absorb relic neutrinos like $ ^{187}Re $ and the 
other half has the same weight, but cannot absorb relic neutrinos. The `neutrino wind' exerts a torque to this cylindrical object and turns 
it so into the `neutrino wind', so that the torque is zero.

\item  The search for the  $C \nu B$ with the induced beta decay~\cite{Faessler4,Lazauskas}. 
\begin{equation}
 \nu_e + ^3H  \rightarrow  ^3He + e^- .
\label{induced}
\end{equation} 
The  signal would show up by a peak in the electron spectrum with an energy of  the neutrino mass above the Q value. This reaction has no threshold and therefore should take place in the environment of  the low energy $C \nu B$.

\end {itemize}

In the present contribution we focus on the last approach based on the $C \nu B$ with the induced beta decay.
For the estimate average relic neutrino density of $ 56 $ neutrinos 
per $ cm^3 $ observation of this process with the existing technologies is not realistic.   
The chances may be improved for the smaller mass of the relic electron electron neutrinos, with the 
increase of the relic neutrino density due to gravitational 
clustering and  by the strength of the beta decay source (specifically the intensity of the tritium source. 
In chapter 4 after eq. $ (\ref{RateKK})$ we discuss this possibility, which would allow tp find the relic neutrinos.).   
Despite of this pessimistic situation with the detection of  $C \nu B$,  from the searches of the reaction  (\ref{induced})  
at least  it should be possible to derive an upper limit for the overdensity of the relic neutrinos 
in our galaxy due to gravitational clustering.  We shell comment this point in the present note. 

In sec. 2 and 3 we briefly give a cosmological background relevant for the subject in question with motivations of 
searching  for the $C \nu B$. In sec. 4 we present a formalism for analysis of both the relic neutrino capture 
reaction (\ref{induced}) and  tritium beta decay. Then we discuss prospects of searches for  the reaction (\ref{induced}) 
with the KATRIN experiment ~\cite{Drexlin}.

\vspace{1cm}

\section{Cosmology of the Microwave Background}

\vspace{0.5cm}

The detection and measurement of the Cosmic Background Radiation in 1964 by Penzias and Wilson $~\cite{Penzias} $ 
surprised, because the radiation followed 
very accurately Planck's black body radiation law:

\begin{equation}
\epsilon(f) df =  \frac{8 \pi h}{c^3} \cdot \frac{f^3 df}{ exp(h f/k_B T_0) - 1} \:  [Energy/Volume] 
 \label{Pl}
\end{equation}

The temperature parameter $ T_0 $ fitted by $ (\ref{Pl}) $ is independent of the observation direction up to the fourth digit. 
The average temperature of the photons in the CMB is about $ 3\cdot T_0 $.

\begin{equation}
T_{0,rad} \equiv T_{0,\gamma} = 2.7255 \pm 0.0006  \: [Kelvin]
\label{Temp}
\end{equation}

Integration over all frequencies $ \it f $ yields the Stefan-Boltzmann law for the energy density. 

\begin{equation}
\epsilon_{rad}  =  \varrho_{rad}c^2 = \alpha T_0^4; \hspace{1cm} \alpha = \frac{\pi^2 k_B^4}{16\hbar^3 c^3};
\label{SB1}
\end{equation}

With: 

\begin{equation}
 k_B \approx  0.86\cdot 10^{-4} \: [eV/Kelvin]; \hspace{1cm} 1 [eV] \approx 11600 \:  [Kelvin];
\label{SB2}
\end{equation}

The average number of photons can be calculated by dividing $(\ref{SB1}) $ by the energy $ 3 \cdot k_B  T_0 $ of the average photons 
in the CMB.

\begin{equation}
 n_\gamma \approx 370  \: cm^{-3}
\label{ng}
\end{equation}

The size of the `hot spots' above the average temperature of the CMB with $ T_{0\gamma} = 2.7255 [Kelvin] $ 
indicates a flat universe with the critical density:

\begin{equation}
 \varrho_c = \frac{3 H^2}{8 \pi G} = 0.94 \cdot 10^{-26}\:  [kg \cdot m^{-3}]; \hspace{1cm} G_{gravitation} 
= 6.67 \cdot 10^{-11} \: [ m^3 kg^{-1} sec^{-2}];
\label{rhoc}
\end{equation}

The baryon density in the universe can be extracted from the relative abundance of the light nuclear isotopes, including the 25 \% mass fraction 
of $ ^4He $. This  yields a baryon density in units of the critical density $(\ref{rhoc}) $ of:

\begin{equation}
 \Omega_{baryon} = \frac{\varrho_{baryon}}{\varrho_c} = 0.04;
\label{Omegab}
\end{equation}

The photon energy density in the same units is:

\begin{equation}
 \Omega_{rad} = \frac{\varrho_{rad}}{\varrho_c} = 4.90\cdot 10^{-5};
\label{Omegarad}
\end{equation}

Although the energy density of the photons (rad) is much smaller than the one of the baryons, the number density 
of the photons is nine orders larger than the one of the baryons.

\begin{equation}
 \frac{n_\gamma}{n_{baryon}} \approx 1.7\cdot 10^9;
\label{ngamma}
\end{equation}

The photons decouple from matter as soon as the electrons get bound to the protons and to the  $ ^4He $ nuclei 
and the universe is electrically neutral. This does not happen, as one could expect,  
after the temperature of the universe goes below the ionisation energy of hydrogen of 10 to 13 [eV].  
Due to the nine orders of photons more than baryons it is enough, if one photon 
of the about $10^9$ per baryon at the high energy end of the Planck distribution $ ( \ref{Pl}) $ ionizes 
the hydrogen atoms. So the decoupling of the photons in the CMB happens at the lower 
temperature of about $3000 $  Kelvin $\approx 0.3 \: eV $ roughly 380 000 years after the big bang. 
So since about 13 billion years the CMB move freely. 
The sphere of the last photon scattering has a radius of 13 to 14 billion light years around the observer. 

\vspace{1cm}

\section{Cosmology of the Cosmic Neutrino Background}
\vspace{0.5cm}

The neutrinos are fermions and thus the Planck distribution (\ref{Pl}) is modified accordingly.  

\begin{equation}
\epsilon(f) df \propto  \frac{f^3 df}{ exp(h f/k_B T_0) + 1} [Energy/Volume] 
 \label{PlB}
\end{equation}

Again { \it f }  is the frequency and { \it h } Planck's Wirkungsquantum.  

The integration over the frequencies yields the total energy density of the neutrinos. Division by the average energy of relic neutrinos  
$3\cdot k_B T_0 = 3\cdot k_B \cdot (T_0 = 1.95\: Kelvin  \: (today)) $ yields the average number density for electron-neutrinos.

\begin{equation}
n(today)_{\nu, e} = \frac{\epsilon(today)_{\nu,e}}{ 3 \cdot k_B \cdot (T_0 = 1.95 \: Kelvin)} = 56 \: cm^{-3}
 \label{nn}
\end{equation}

For the total neutrino density this number has to be multiplied by 6. 

\vspace{0.5cm}
 
The neutrinos decouple much earlier from matter than the photons due to their very weak interaction. The competition for decoupling of the neutrinos is 
between the expansion rate of the universe given by the Hubble constant

\begin{equation}
 H = \frac{\dot{a}}{a} = \sqrt{\frac{8 \pi G}{3} \varrho_{total}} = \sqrt{\frac{8 \pi \varrho}{3 M_{Planck}^2}} \propto a^{-4};   
\label{Hubble}
\end{equation}
and the reaction rate for relativistic neutrinos:

\begin{equation}
 \Gamma = n_\nu<\sigma v> \approx T_0^3G_{Fermi} T_0^2= G_F T_0^5 \propto a^{-5} ; \hspace{1cm} v \approx c=1 .
\label{neutrinoR}
\end{equation}
The above equations are given in the natural units $ \hbar  = c = 1 $ and $ a \propto 1/T $ is the scale parameter of the universe. 
For decreasing  temperature T and increasing scale a 
neutrino reaction rate and the neutrinos decouple, when the Hubble expansion rate $ (\ref{Hubble}) $ 
is about equal to the neutrino reaction rate $(\ref{neutrinoR})$: 
\begin{eqnarray}
\mbox{in radiation dominated era: } \quad T_0  \: = \bigg(\frac{45 \hbar^3 c^5}
{32\pi^3 G}\bigg)^{1/4}\cdot \frac{1}{ t^{1/2}} \equiv  1.3\bigg(\frac{t}{sec} \bigg)^{-1/2} \: [MeV]; \nonumber
\label{DecouplingN}
\end{eqnarray}
This corresponds to the time after the Big Bang of about $1 \ second $.
\vspace{0.5cm}

To find the temperature of the neutrino background today one has to take into account, that the photon background is heated 
up by the positron-electron anihilation into two photons after the decoupling of the neutrinos. This anihilation happened with constant entropy.

\begin{equation}
 e^- + e^+ \rightarrow \gamma + \gamma ; \hspace{1cm} S \propto g_i \cdot T_{0i}^3 = g_f \cdot T_{0f}^3 = const;
\label{electronA}
\end{equation}

Constant entropy means equal degrees of freedom. The initial state of this phase transition are the electrons $(g_{e^-} = 2 \cdot 7/8)$, 
positrons $(g_{e^+} = 2 \cdot 7/8) $ and photons $(g_\gamma = 2)$. While in the final state one has the photons only $(g_\gamma = 2)$. 
The factor 7/8 takes into account the Pauli principle  for the electrons. The temperature dependence $ T_{i/f}^3 $  
is proportional to the degrees of freedom in momentum space. 

\begin{eqnarray}
g_{i} = 4 \cdot \frac{7}{8} + 2 
= \frac{11}{2};
 \hspace{0.5cm} g_{f} & = & 2; \hspace{0,5cm} \frac{g_{f}}{g_{i}} = \frac{4}{11} =
 \bigg(\frac{T_{0,i=\nu}}{T_{0,f=\gamma} = 2.725} \bigg)^3; \\
 T_{0,\nu} (today) & = & 1.95 \:  [Kelvin]; \nonumber
\label{Entropy}
\end{eqnarray} 

The relativistic massless neutrino energy density in relation to the critical density is now expressed 
by the the photon background $ (\ref{Omegarad}) $ given by:

\begin{equation}
\Omega_\nu = 3 \cdot \frac{7}{8} \cdot \bigg(\frac{4}{11}\bigg)^{4/3} \cdot \Omega_{rad} = 0.68 \cdot \Omega_{rad}; 
\label{OmegaN}
\end{equation} 

The factor 3 originates from six different types of neutrinos and two photon degrees of freedom ($ 6/2$). 
Today the neutrinos are non-relativistic due to their finite mass and the low kinetic energy of about
 $3 \cdot T_0 = 3 \cdot 1.95[Kelvin] \equiv 5.5\cdot 10^{-4} [eV]$.
For massive neutrinos the energy density in units of the critical density is:

\begin{equation}
\Omega(m_\nu \neq 0)_\nu =  \Omega( m_\nu = 0) \cdot \frac{\sum m_\nu c^2}{3 \cdot k_B T} = \frac{\sum m_\nu c^2}{ 45 eV}; 
\label{OmegaM}
\end{equation} 

In summary the cosmological development starts with the radiation dominated era from the Big Bang (BB) till about 30 0000 years
after the BB with a temperature of $ 1 \: eV \equiv 10 \, 000 \:Kelvin $ at that time. In this era the neutrinos decouple at about  1 second after the 
BB and $ T \approx 10^{10}$ Kelvin. The energy density scales with  $a^{-4} \propto T^4$. The matter dominated era lasts from about 30 000 
to $ 8\cdot 10^9 $ years after BB. In this era the photon background decouples at around 3000 Kelvin and 300 000 years after the BB. 
The energy scales during this time with $ a^{-3} \propto T^3 $. At around $ 8\cdot 10^9 $ 
years after BB the dark energy takes over with a constant 
energy density, which accelerates the expansion. 
\vspace{1cm}
 
\section{Search for the Cosmic Neutrino Background with KATRIN.}
\vspace{0.5cm}

Here we discuss the relic neutrino induced beta capture $(\ref{induced})$ and the beta decay of tritium 
\begin{equation}
\label{decayH}
^3H \rightarrow ^3He + e^-  + \bar{ \nu}_e    ;   \hspace{1cm}    Q = 18.562 \:  keV  
\end{equation} 
with the focus on the
KATRIN experiment $~\cite{Drexlin} $ which we believe has the detection potential to set an upper limit for the relic electron neutrino 
density in our neighborhood. 

The electron spectrum of the reactions $(\ref{induced}) $ and  $ (\ref{decayH})$ 
for tritium has for infinite good energy resolution the form:

\begin{eqnarray}
\frac{dN_e}{dE} = K \cdot F(E;Z) \cdot p_e E_e (E_0 - E_e) \sum_{j=1}^3 |U_{ej}|^2 \sqrt{(E_0 - E_e)^2 -m(\nu_j)^2}  \nonumber\\ 
+ \eta \cdot n_{\nu,e}\cdot \delta\bigg(E - Q -\sqrt{\sum_{j=1}^3 |U_{ej}|^2 m(\nu_j)^2}\bigg); 
\label{Spectrum}
\end{eqnarray}

with:

\begin{equation}
E_0 = Q + m_e; \hspace{0.5cm}  Q = 18.562 \: keV; \hspace{0.5cm} E_e = \sqrt{m_e^2 + p_e^2}; \hspace{0.5cm} E = T_e = E_e - m_e; 
\label{SpectrumE}
\end{equation} 
and the flavor state $\nu_e $ expressed by mass eigenstates $ \nu_j$:
 
\begin{equation}
\nu_e = \sum_{j=1,2,3} U_{ej} \nu_j  ; \hspace{1cm} m_{\nu,e}^2 = \sum_{j=1,2,3} |U_{ej}|^2 m_{\nu,j}^2;
\label{neutrinoM}
\end{equation} 

To determine the neutrino mass one must fit the electron spectrum at the upper end of the Kurie plot of the tritium decay  $(\ref{decayH}) $ 
the Q value with an accuracy of the order of milli-volt (mV) and the electron neutrino mass $(\ref{neutrinoM})$. To determine the local 
relic neutrino density or an upper limit for it, one has in addition to fit $\eta \cdot n_ {\nu,e}$ from eq. $( \ref{Spectrum})$, which is 
proportional to the local relic electron-neutrino density $ n_{\nu,e} $  (see below). 

The tritium beta decay probability is according to Fermi's Golden Rule given by:

\begin{equation}
\Gamma_{decay}^\beta(^3H) = \frac{1}{2 \pi^3}\cdot \sum \int |<^3He|T|^3H>|^2 \cdot 2\pi \delta(E_\nu+E_e + E_f - E_i)
 \frac{d\vec{p}_e}{2\pi^3}\cdot \frac{d\vec{p}_\nu}{2\pi^3};
\label{tritium}
\end{equation} 
One averages over the initial and sums over the final magnetic substates and integrates over the electron and neutrino momenta.
The Beta decay matrix element contains Fermi and Gamow-Teller contributions.

\begin{eqnarray} 
\sum |<^3He|T|^3H>|^2 & =  &2 \cdot (\frac{G_F \cos(\vartheta_C)}{\sqrt{2}} )^2 \cdot F_0(Z+1,T_e) [B_F(^3H) + B_{GT}(^3H)];
 \nonumber\\
	B_F(^3H) & = & |M_{Fermi}|^2 = \frac{1}{2}|<He(1/2)^+||\sum_n \tau_n^+||^3H(1/2)^+>|^2;
\\
B_{GT}(^3H) & = & |M_{Gamow-Teller}|^2 = \frac{1}{2}|<He(1/2)^+||\sum_n \tau_n^+  \sigma_n||^3H(1/2)^+>|^2;\nonumber 
\label{ReducedB}
\end{eqnarray}
$F_0(Z+1, T_e) $ 
is the  Fermi function $~\cite{Doi85}$, which takes into account the Coulomb distortion 
of the outgoing s-electron with the asymptotic kinetic energy $ T_e $
in the final nucleus with the charge $Z+1 $. The matrix elements can be calculated analytically by angular momentum and isospin algebra. 
By integrating over the phase space one obtains:

\begin{eqnarray} 
\Gamma_{decay}^\beta(^3H) = \frac{1}{2\pi^3} m_e(G_F \cos(\vartheta_C) m_e^2)^2 [B_F(^3H) + B_{GT}(^3H)] \cdot I^\beta(^3H);
 \nonumber\\
B_F(^3H) + B_{GT}(^3H) = 5.645; \hspace{1cm} I^\beta(^3H) = 2.88 \cdot 10^{-6}; 
\label{ReducedBB}
\end{eqnarray}

The half life is then given by:

\begin{equation}
Theory: \hspace{0.5cm} T^\beta_{1/2} = \frac{\ln2}{\Gamma_{decay}^\beta(^3H)} = 12.32 \hspace{0.2cm} years; 
\hspace{1cm} Experiment: T^\beta_{1/2} = 12.33 \hspace{0.2cm} years ;
\label{Thalf}
\end{equation} 

The same reduced Fermi and GT transitions are needed for the determination of the induced relic 
neutrino capture reaction $( \ref{induced})$. 

\begin{eqnarray} 
\Gamma_{capture}^\beta (^3H)  =  \frac{1}{\pi} (G_F \cos(\vartheta_C))^2 F_0(Z+1, T_e)
[B_F(^3H) + B_{GT}(^3H)] p_e T_e \cdot <n_{\nu,e}> \frac{n_{\nu,e}}{<n_{\nu,e}>}  \nonumber\\
  =  4.2 \cdot 10^{-25} \frac{n_{\nu,e}}{<n_{\nu,e}>}[\mbox{for 1 tritium atom/year}]; \hspace{4cm}\nonumber \\  
	with: \hspace{0.3cm} 
<n_{\nu,e}> \:  = 56 \hspace{0.2cm} cm^{-3}; \hspace{6cm} 
\label{RateN}
\end{eqnarray}

The expression $(\ref{RateN})$ gives the relic neutrino capture rate per year by  only one tritum atom. 
It is proportional to the local overdensity $ n_{\nu,e}/<n_{\nu,e}> $ 
with  to today's average relic neutrino density $ <n_{\nu,e}> \:  \approx  56 \hspace{.2cm} $
[electron-neutrinos $ / cm^3] $ for $ T_0 = 1.95 $ Kelvin in the universe. 

The effective mass of the Tritium source of KATRIN corresponds to this part of the source activity, of which the decay 
electrons arrive without inelastic scattering (and energy loss) in the detector for counting. 
Different values are given in the literature for 
this effective mass. Kaboth et al. $ ~\cite{Kaboth} $ give $ 66 \: microgram $.  In a previous publication
$ ~\cite{Faessler4}$  we assumed after consultation with the KATRIN collaboration  $ 50 \:  \mu g $. Guido Drexlin 
$ ~\cite{Drexlin2} $  told 
us, after we finished this manuscript, that the correct value is $ 20 \: \mu g $. Thus we modified our results for the 
effective mass to the value $ 20\: \mu g $.  This means $ 2\cdot10^{18} \quad  Tritium_2 $ molecules. The capture rate 
of relic neutrinos is then:

\begin{equation}
\mbox{Capture rate at KATRIN: } N_\nu (KATRIN) = 1.7\cdot 10^{-6} \cdot \frac{n_{\nu,e}}{<n_{\nu,e}>}; 
\label{RateK}
\end{equation} 

For the average relic neutrino number density $ <n_{\nu,e}> \: = \: 56 \: cm^{-3} $ this corresponds to every 590 000 years a count. 
So the critical question is how much is the number density of the relic neutrinos increased by gravitational clustering in the solar system or better in our galaxy.
Due to the small mass of the neutrinos one expects a large free streaming length and thus a clustering on the scale of galaxies and their halo of about 1 Mpc or even of galaxy clusters 
around 50 Mpc. 

Ringwald and Wong $~\cite{Ringwald}$ investigate the relic neutrino gravitational clustering as 
a byproduct of the clustering of dark matter. They start with density 
profiles of dark matter for different virial-masses (from $10^{11} \: to \:  10^{16} $ solar masses) 
of galaxies  and calculate with the Vlasov equation 
trajectories of the relic neutrinos for different neutrino masses $ m_\nu = 0.15eV;\quad 0.3 eV; \quad 0.6eV $. 
The relic neutrino number overdensities are 
shown in their $~\cite{Ringwald}$ figure 3 and 4 and vary between the value 1 and $10^4$. Ref. $~\cite{Ringwald}$ 
shows also, that the gravitational clustering 
of the neutrinos is possible on the scale of galaxies and their halos of about 1 Mpc. 

Lazauskas, Vogel and Volpe $~\cite{Lazauskas}$ assume, that the neutrino overdensity is proportional to the 
baryon overdensity in galaxy clusters.
(Average number baryon density in the universe: $ <n_b> = 0.22\cdot 10^{-6}\hspace{0.3cm} cm^{-3} $.)  
They estimate a neutrino 
overdensity of $ n_{\nu,e} = 10^3$  to  $10^4 $ on the scale of 50 Mpc for galaxy clusters. If one 
assumes the result of Ringwald and Wong $~\cite{Ringwald}$,
that relic neutrinos can cluster on the scale of a single galaxy and their halo  and uses 
the proportionality to the baryon overdensity of Lazauskas et al.$~\cite{Lazauskas}$,
then one can expect very optimistically overdensities up to 
$ n_{\nu,e}/<n_{\nu,e}> \: \le  10^6 $ in our neighbourhood. With this optimistic estimate of the  upper limit 
for the relic neutrino overdensity of $ 10^6$ one obtains from 
equation $(\ref{RateK})$ :
 
\begin{equation}
 N_\nu (KATRIN) = 1.7\cdot 10^{-6} \cdot \frac{n_{\nu,e}}{<n_{\nu,e}>} \: [year^{-1}] \approx 1.7 \: [counts\:  per\: year] ; 
\label{RateKK}
\end{equation} 

This seems not possible to measure for the moment. One way out would be to increase the effective 
activity of the tritium source. An effective mass of 2 milligrams Tritium  would mean with the above optimistic estimate of the 
relic neutrino number overdensity $ n_{\nu,e}/<n_{\nu,e}> \:  \approx  10^6 $ about $170$ counts per year, 
which should be feasible. 
\newline
The possibility to increase the the Tritium source strength can be seen from 
figure 15 of the KATRIN Design Report $~\cite{Design}$. 
The increase of the tritium source strength is limited by the scattering of the emitted electrons by the tritium gas. 
After the mean free path

\begin{equation}
\lambda_{mean\  free \ path} = \frac{1}{\rho \cdot \sigma(electron-tritium)} = d_{free}
\label{mean}
\end{equation}

only about $ 37 \% $ decay electrons have not yet scattered. All the others including also electrons with the maximum energy, 
which contain the information on the neutrino mass and on the relic neutrino capture by tritium, are lost for the measurement. 
A detailed analysis $~\cite{Design}$ 
shows, that the maximum number of unscattered decay electrons can escape the source within the last area 
of the tritium gas of a width of half the mean free path. 
Thus to increase the column density
 (number of tritium atoms in a column with the base area $1 cm^2 $ and a specific length d of the column) 
of the tritium gas source beyond this value 
does not increase the the effective source strength, but yields only more background. Thus an increase of the tritium source 
strength with the present geometry a source with the area of about $ 50 \  cm^2; \  8\  cm $ \  in diameter is not able 
to increase the source strength. 
The increase of the source area to $5000 \  cm^2, \  80 \  cm \  diameter$ looks like a way out. 
But also this seems to be impossible. 
\newline
The magnetic flux at the present source is given by

\begin{equation}
Magnetic Flux(source) = 53\ cm^2 \cdot 3.6 \ Tesla \approx 190\  Tesla \cdot cm^2,
\label{flux1}
\end{equation}

which must be the same in the spectrometer.

\begin{equation}
Magnetic Flux(spectrometer) = 63.6\ m^2 \cdot 3 \ Gauss \approx 190\  Tesla \cdot cm^2,
\label{flux2}
\end{equation}

The low magnetic field in the spectrometer of only $3\ Gauss$ is needed to transform the electron momenta, 
which are at the source almost perpendicular to the beam direction due to the cyclotron motion, 
into almost a translational direction (see figure 9 of the KATRIN design report  $~\cite{Design}$), 
to reject with an electric opposing field all electrons 
almost up to the tritium Q-value ($Q = 18.562\  keV $).  
To increase the source strength by a factor $ 100 $ with a corresponding increase of the source area 
to $ 5000 \ cm^2, ~80\ cm $  diameter, one needs also to increase the spectrometer cross section by a 
factor $100$ or the diameter to about $90 \ m$, which is not possible. 
\newline
If one increases the source area to $5000 \ cm^2$ and thus the diameter to about $ 80\ cm$ and reduces 
at the same time the magnetic field at the source to  $ 360 \  Gauss$, one can conserve the magnetic flux 
with the $3 \ Gauss$ in the present spectrometer. But an $80 \  cm$ wide beam does not fit in the existing  
transport channel. So one needs a strong magnetic field of about $3.6 \  Tesla$ to compress the 
electron (cyclotron) beam to about $8 \ cm $ for the existing transport channel. But increasing the magnetic 
field would reverse the electron beam by the magnetic mirror effect. Thus one needs to accelerate the 
electron beam to overcome the magnetic mirror. This means to put the whole transport channel to a high voltage. 
Before entering the spectrometer the beam must be again on earth potential. This manipulation would allow 
to use the existing spectrometer with a $3 \ Gauss$ field. 
\newline

Till now we did not include the requirement for an energy resolution of the spectrometer  of about 
$\Delta E\:  \approx \: 1.0 \: eV $  (see ref. $~\cite{Drexlin}$). 
The energy resolution of a KATRIN type spectrometer is determined by the perpendicular energy of the decay 
electrons  in the spectrometer in the cyclotron motion  $E_{f\perp}$. 
The electrons  with a relatively large  perpendicular energy in the spectrometer with longitudinal 
energy just below the Q-value cant be rejected by the opposing electric field and therefore 
arrive all in the detector. The angular momentum in the circular cyclotron motion 
of an electron and also the corresponding magnetic moment of this ring current conserve the ratio of the 
perpendicular energy over the magnetic field of the electrons:

\begin{equation}
 |\vec{L}| = |\vec{r}\times \vec{p}| \propto \mu = const \propto \frac{E_{i\perp}}{B_i} = \frac{E_{f\perp}}{B_f}
\label{moment}
\end{equation}

An energy resolution of about $ \Delta E \ = \ 1\  eV $ thus requires:

\begin{equation}
 \Delta E = 1 \ eV\ = \ E_{f\perp}\ = \  \frac{B_f}{B_i} \cdot E_{i\perp} \ = \ \frac{ 3 \ Gauss}{360 \ Gauss} E_{i\perp}
\label{resolution}
\end{equation}

Thus at the Tritium source one can have only $ E_{i\perp}\  = \ 120 \  eV $ in the perpendicular motion. 
The rest of the Q-value of $ 18.5 \ keV $ must be in the longitudinal direction at the source. 
(Only the momentum of the electrons is a vector and can be decomposed in a 
longitudinal part and a transversal part. The longitudinal and the transversal energy do not add up to the Q-value.) 
This allows only a cone with a small opening of $\vartheta_i \  = \ 5.7^{\circ } $ relative to the beam axis. 
The accepted space angle of the emitted electrons is therefore reduced to 
$\Delta \Omega/(2\cdot\pi) \  = \  0.005 \rightarrow 0.5 \ \% $.
Thus this requirement for the energy resolution reduces the electron beam intensity by a factor $1/200$. 
As a whole with an increase of $100$ from the size of the source and a decrease of $1/200$ 
due to the required energy resolution one looses intensity. 
\newline 
With the KATRIN spectrometer and the resolution of $\Delta E \ = \ 0.93 \  eV $ and the background  
one hopes to reduce the upper limit of the electron 
neutrino mass to about $ 0.2 eV \  (90 \% \  C.  L. )$.
Fitting at the upper end of the Kurie plot at $Q - m_{\nu,e}$ the electron 
spectrum $(\ref{Spectrum})$ the KATRIN collaboration hopes to determine the Q-value and the neutrino mass. 
The electron  peak due to the capture of the relic neutrinos lies at $ Q +m_{\nu,e} $ .
The neutrino mass and the energy resolution and the background remain the same as for the determination of the neutrino mass. 
One has only one additional fit parameter more (or two, if one counts the width of this peak,), 
the counts in the peak at $Q + m_{\nu,e} $. 
At the moment it does not seem possible to detect with a KATRIN type spectrometer 
the Cosmic Neutrino Background. But one should be able to give an upper limit for the local 
relic neutrino overdensity $ n_{\nu,e}/<n_{\nu,e}> $ in our Galaxy. 
\vspace{1cm}

\section{Conclusions.}

\vspace{0.5cm}

The Cosmic Microwave Background (CMB) of photons allows to look back in our universe 
to about 380 000 years after the Big Bang (BB). Today's CMB
 has a temperature of 
$ T_0 = 2.7255\pm 0.0006 \: [Kelvin] $. The Cosmic Neutrino Background decoupled 
much earlier about 1 second after the BB and has today a 
temperature of $ T_0 \, = \, 1.95 \;[Kelvin] $. Thus one can look back to about 1 second after the BB, if one is 
able to measure details of the Cosmic (relic) Neutrino Background. 
\newline
There are two major problems, which make the detection of the neutrino 
background very difficult and perhaps at least today impossible:

\begin{itemize}

\item With the average relic electron neutrino number density of $ <n_{\nu,e}> \:  = 56 \: cm^{-3}$ KATRIN 
could measure only every 590 000 years a count. So the hope is with the local 
overdensity due to
gravitational clustering of the neutrinos in our galaxy. Estimates for this overdensity  
$ n_{\nu,e}/<n_{\nu,e}> $  vary widely from about $10^2$ to $10^6$. 
With the optimistic estimate of a local overdensity of $ 10^6$ one obtains with KATRIN 1.7 counts per year. 
If one could increase the effective mass of the tritium source from 
20 micrograms to 2 milligrams, this optimistic estimate of the overdensity would mean 170 counts per year. 

\item The second problem could be the energy resolution of the KATRIN spectrometer of $\Delta E = 0.93 \quad  eV$.
With this resolution one expects to 
extract from  the electron spectrum $(\ref{Spectrum})$ at the upper end of the Kurie plot at $ Q - m_{\nu,e} $
a very accurate Q-value of the accuracy  of milli-eV and an upper limit of the electron neutrino mass of 
about $ m_{\nu,e} \le  0.2 \quad [eV]  \quad 90 \% \quad C.L. $ with an  
energy resolution of $ 0.93 \: eV$  and the background. 
If one can fit the Q value and the electron neutrino mass accurately enough, the position of the electron   
peak from the induced capture of the relic neutrinos is known to be at an electron energy of $ Q\, + \,  m_{\nu,e} $. 
The energy resolution and the background is the same at $ Q - m_{\nu,e} $ and at $ Q + m_{\nu,e} $. One should with KATRIN at 
least be able to determine an upper limit for the local relic neutrino density.

\end{itemize}
\vspace{0.5cm}

In chapter four we discussed critically methods how to increase the strength 
of the tritium source from 20 micrograms to 2 milligrams  
with the existing spectrometer. Such an increase of the tritium source could yield 170 counts from the capture of relic 
(Cosmic Background) neutrinos. Presently such an increase seems not possible. 
But one should be able to find an upper limit for the local overdensity $ n_{\nu,e}/<n_{\nu,e}> $  
of the relic neutrinos (Cosmic Neutrino Background) in our galaxy.

\vspace{1.5cm}

Acknowledgments:

\vspace{0.5cm}

We want to thank Guido Drexlin and Christian Weinheimer for discussions about the 
feasibility to increase the source strength of the KATRIN spectrometer. 
This work was supported in part by the Deutsche Forschungsgemeinschaft
within the project "Nuclear matrix elements of Neutrino Physics and Cosmology"
FA67/40-1 and  by the grant of the Ministry of Education
and Science of the Russian Federation (contract 12.741.12.0150).
F. S. acknowledges the support by the VEGA Grant agency
of the Slovak Republic under the contract No. 1/0876/12 and by the Ministry 
of Education, Youth and Sports of the Czech Republic under contract LM2011027.
S. K. thanks for support by the FONDECYT grant 1100582 and the Centro Cientifico-Tecnol\'{o}gico de 
Valpara\'{i}so PBCT ACT-028.

\end{document}